# Abrupt Climate Transition of Icy Worlds from Snowball to Moist or Runaway Greenhouse

Jun Yang[1,*], Feng Ding[2], Ramses M. Ramirez[3], W. R. Peltier[4], Yongyun Hu[1,*], and Yonggang Liu[1]

[1] Laboratory for Climate and Atmosphere-Ocean Studies, Department of Atmospheric and Oceanic Sciences, School of Physics, Peking University, Beijing, China.
[2] Department of the Geophysical Sciences, University of Chicago, Chicago, IL, USA.
[3] Carl Sagan Institute, Department of Astronomy, Cornell University, Ithaca, NY, USA.
[4] Department of Physics, University of Toronto, Ontario, Canada.

*Correspondence to J.Y., junyang@pku.edu.cn and Y.H., yyhu@pku.edu.cn

**Abstract**: Ongoing and future space missions aim to identify potentially habitable planets in our Solar System and beyond. Planetary habitability is determined not only by a planet's current stellar insolation and atmospheric properties, but also by the evolutionary history of its climate. It has been suggested that icy planets and moons become habitable after their initial ice shield melts as their host stars brighten. Here we show from global climate model simulations that a habitable state is not achieved in the climatic evolution of those icy planets and moons that possess an inactive carbonate-silicate cycle and low concentrations of greenhouse gases. Examples for such planetary bodies are the icy moons Europa and Enceladus, and certain icy exoplanets orbiting G and F stars. We find that the stellar fluxes that are required to overcome a planet's initial snowball state are so large that they lead to significant water loss and preclude a habitable planet. Specifically, they exceed the moist greenhouse limit, at which water vapour accumulates at high altitudes where it can readily escape, or the runaway greenhouse limit, at which the strength of the greenhouse increases until the oceans boil away. We suggest that some icy planetary bodies may transition directly to a moist or runaway greenhouse without passing through a habitable Earth-like state.

Icy worlds are common in the solar system (such as Europa[1], Enceladus[2], Ganymede[3], and early Earth[4]) and plausibly also in extra-solar systems[5]. A fundamental question is that: Would such icy planets and moons become habitable once their ice cover melts? There are two ways for the icy worlds to escape the globally ice-covered snowball states. One is that continuous atmospheric accumulation of $CO_2$ from volcanic outgassing during the snowball phase triggers the melting[6,7]; this is plausible for planets having an active carbon cycle (e.g., Earth), and they become habitable for life after the ice melts. The other is that the stars brighten with time and the ice melts once the stellar flux exceeds a critical value[8]; this is the case for planets and moons lacking an active carbon-silicate cycle and having low concentrations of greenhouse gases (e.g., Europa). Here, we investigate the second case using a series of three-dimensional (3D) climate model experiments. In contrast to previous studies[9-11] that suggest the existence of a habitable world after the snowball deglaciation, we show that the increased stellar insolation will force the planet into an uninhabitable moist or even runaway greenhouse state.

Using 0D and 1D energy balance climate models and a 3D gray-gas atmospheric general circulation model (GCM), previous studies[12-14] had examined the climate evolution of a snowball planet having low concentrations of greenhouse gases. However, their models were unable to account for the effects of clouds, lapse rate, spatial snow and ice distributions, realistic atmospheric radiative transfer, and/or meridional atmospheric heat transport, which have been identified[15,16] to be critical for simulating the snowball climate. Moreover, the simple climate models cannot simulate vertical water vapor transports or the onset of a moist greenhouse state. These studies[12-14] showed that a post-snowball climate should be hot, but whether it is habitable for life or not remains unclear.

**Stellar flux threshold for snowball melting**
First, we calculate the stellar flux threshold required to trigger the melting of a snowball planet orbiting different types of stars (from F to K), using the 3D global GCM CAM3 (see Methods). Model comparisons[17-19] had demonstrated that CAM3 could accurately simulate a snowball Earth climate that possibly occurred in 600-800 million years ago. In the second step, we fix the insolation at the stellar flux threshold and allow the ice to melt, in order to examine whether the post-snowball planet will fall into a moist or runaway greenhouse state. The moist greenhouse state is defined as that in which water vapor concentration in the upper atmosphere is high enough that $H_2O$ photo-dissociation and subsequent H escape to the space becomes significant[8,20]. When the water vapour mixing ratio reaches $3 \times 10^{-3}$, a planet having an equivalent water inventory of Earth's would lose all of the water within the present-day age of the Solar System, 4.5 billion years[8]. The runaway greenhouse state is defined as that in which the atmosphere becomes optically thick at all infrared wavelengths due to the water vapor greenhouse effect, absorbed stellar flux exceeds allowed maximum outgoing longwave radiation, and surface liquid water evaporates entirely into the atmosphere[12,21]. These two climate states constrain the inner edge of surface liquid water habitable zone[8].

For a solar spectrum, a fixed $CO_2$ concentration of 300 ppmv (i.e., close to the pre-industrial level), an ice albedo of 0.5 (the typical albedo of sea ice on Earth[22]) and a snow albedo of 0.8, the stellar flux required to trigger the snowball melting is 1,800 W m$^{-2}$ (Fig. 1), which is 32% higher than the solar flux for the modern Earth. In the snowball state, the surface is cold, mainly because the surface albedo is high[12,13] and the greenhouse effect is weak (4-14 K in global mean, versus 33 K for Earth) due to low water vapor concentration[15]. For instance, in the case of a solar flux of 1,400 W m$^{-2}$, the planetary albedo is 0.71 (Europa's albedo is ~0.62[1] and Enceladus's ~0.90[2]), the vertically integrated water vapor amount is only 0.2 kg m$^{-2}$ (or 1% of modern Earth's), and the global-mean surface temperature is only 214 K. As the stellar flux is increased, the climate sensitivity is 0.2-0.8 K per W m$^{-2}$, much lower than that of the modern Earth[23], 0.7-1.9 K per W m$^{-2}$. Clouds have a small shortwave cooling effect due to the masking of the high surface albedo and a moderate longwave warming effect (7-20 W m$^{-2}$ in global mean). In the region around 15 °S(N), the planetary albedo is lower than that in other latitudes (Fig. 1b), which is due to the fact that there is nearly no snow over ice in this region where sublimation is stronger than snowfall. When the surface is set to have a surface albedo of 0.6 everywhere (that of sea glacier, which is compressed thick snow overlying ocean[24]), the stellar flux threshold is still high, 1,600 W m$^{-2}$ (Fig. S1).

Sensitivity tests show that the stellar flux threshold is around 1,500-1,900 W m$^{-2}$ for a wide range of parameters, including ice cloud particle size, planetary obliquity, eccentricity, gravity, radius, background gas concentration ($N_2$), and surface topographic height (Fig. 2a). The robustness of our conclusion is due to the simplicity of the mechanism: Any planet that has a high albedo and a low greenhouse gas concentration should have low surface temperature and thereby require a high stellar flux to trigger the melting. When the gravity and radius are set to those of Europa, the stellar flux threshold is still high, 1,550-1,600 W m$^{-2}$. Increasing the N$_2$ partial pressure from 1 to 10 bars has a small (< 3 K) effect on the maximum surface temperature and thereby nearly does not affect the stellar flux threshold. This is due to the compensation between the warming effect through increasing pressure broadening and the cooling effect via enhancing Rayleigh scattering and increasing equator-to-pole atmospheric heat transport (Fig. S2). Increasing the obliquity from 0º to 30º raises the stellar flux threshold by 100-300 W m$^{-2}$ (depending on the surface albedo), primarily due to the fact that less solar energy is deposited at the equator, where the annual-mean surface temperature is the maximum. Halving or doubling ice cloud particle sizes has a very small effect (< 50 W m$^{-2}$) on the result.

Compared to a G-star spectrum, the stellar flux threshold for a F-star spectrum is higher, 1,700-2,200 W m$^{-2}$ (Fig. 2b). This is due to a blue shift of the spectrum[8] and therefore a higher surface albedo[25]. For a K-star spectrum, the stellar flux threshold is lower, 1,150-1,300 W m$^{-2}$ (Fig. 2c) due to the redder spectrum. For G- and F-star systems, the stellar flux thresholds are close to or even higher than both the moist[26-28] and runaway[29-31] greenhouse limits (Fig. 2a). Exceptions are that for slowly rotating planets[26], synchronously rotating planets around M stars[32], and desert planets with limited surface water[33], their moist and runaway greenhouse limits could be higher than the stellar flux thresholds found here.

**Transition to a moist or runaway greenhouse state**
For a G- or F-star system, when the stellar flux deposited on an icy planet reaches the snowball-melting stellar flux threshold, surface ice melts completely and the atmosphere enters into a moist greenhouse state. After the ice melts, surface albedo decreases from 0.5-0.8 to as low as 0.09, total water vapor amount reaches > 500 kg m$^{-2}$ (implying a strong greenhouse effect), and upper-atmospheric water vapor concentration becomes higher than the moist greenhouse level of $3\times10^{-3}$ (Figs. 3 & S3). Energy balance analyses (Fig. 3e) suggest that equilibrium surface temperatures would be 7-19 K higher (for the G-star case, assuming the climate sensitivity[23] is 0.7-1.9 K per W m$^{-2}$) and 35-95 K higher (for the F-star case) than the last converged solutions shown in the figure, if the model could be integrated longer. Decreasing cloud particle size, increasing planetary radius, or reducing gravity makes the planet enter a moist greenhouse state at lower stellar fluxes, and vice versa (Figs. S4 & S5). The most sensitive parameter is the gravity, due to the fact that low gravity decreases atmospheric effective emission pressure and temperature[12].

In the G- and F-star experiments, absorbed stellar radiation is higher than allowed maximum outgoing longwave radiation (Fig. 3f). These planets, therefore, would be expected to enter runaway greenhouse states, in which surface temperature would keep increasing to above 1400 K[8], at which point the surface begins to radiate in the visible wavelengths in order to balance the excess stellar energy. In the K-star experiment, the ice line retreats to 50 ºS(N) and the post-snowball surface temperature, 284 K in global mean (Fig. 3b), is similar to the modern Earth's. This is due to the redder spectrum and thereby a lower surface albedo and a lower snowball-melting stellar flux threshold.

The snowball-melting stellar flux threshold is sensitive to surface albedo, and the runaway greenhouse limit depends on gravity. If surface ice were mixed with dust, surface albedo would be low, which can decrease the stellar flux threshold significantly. For a dirty ice surface having an albedo of 0.4, the threshold is only 1,300 W m$^{-2}$ (Fig. 2a). This threshold is lower than the moist and runaway greenhouse limits for Earth-gravity planets, however, it is still high enough to trigger a runaway greenhouse state on moons that have low gravity. For a moon with gravity of 1.3 m s$^{-2}$ (Europa's), the allowed maximum outgoing longwave radiation[12] is only 238 W m$^{-2}$ and the runaway greenhouse limit is 1,190 W m$^{-2}$ (if assuming the post-snowball planetary albedo is 0.2).

**Implications for planetary evolution**
Our results suggest that an icy exoplanet at the equivalent distance of the modern Earth from the Sun would remain in a snowball state for 10$^9$ years or even longer until its received stellar flux evolves to become much higher than the insolation for the modern Earth. Following this, the planet would directly jump to a moist or runaway greenhouse state (Fig. 4). Europa and Enceladus will have no habitable period. They will transit to a moist or runaway greenhouse state when the Sun becomes a red giant in 6-7 billion years, at which time the stellar flux at the location of Europa will reach the snowball-melting threshold. Our results also suggest that a globally frozen planet can remain not only near the outer edge of the liquid water habitable zone[5,34,35] but also near the inner edge or even further inward, especially in F- and G-star systems. Some exoplanets that were thought to be potentially habitable or to be in a runaway greenhouse state (such as Kepler-22b[36] around a G star, which receives ~1.1 times Earth's insolation) may actually remain in snowball states if they started with globally ice-covered oceans (i.e., the so-called "cold start"[8]).

Finally, we emphasize that our conclusion applies to planets that have an inactive carbonate-silicate cycle and low concentrations of greenhouse gases in G-star and F-star systems. Compared with Earth-mass planets, smaller planets and moons are more likely at risk of the abrupt climate transition because they lose their interior heat faster and volcanic outgassing would cease earlier[37]. An exoplanet more massive than Earth may also be at the risk if its lithosphere is in the stagnant lid regime like on Venus and Mars[38-39]. Our conclusion, however, does not apply to Earth or Earth-like planets that have an active carbonate-silicate cycle. When $CO_2$ concentration is higher, the planet would require a smaller stellar flux to trigger the melting due to the $CO_2$ greenhouse effect (Fig. S6). Under a solar flux of 94% of the present level during the Neoproterozoic era 600-800 million years ago, the $CO_2$ concentration required to melt the snowball Earth is > 100,000 ppmv[15,17,18]. The post-snowball Earth is hot, but it is in neither a runaway nor a moist greenhouse state (Fig. S7-S9). This is because the solar flux is much lower than the runaway greenhouse limit and the high level of $CO_2$ acts to cool the upper atmosphere and thereby decrease the $H_2O$ concentration there[40].

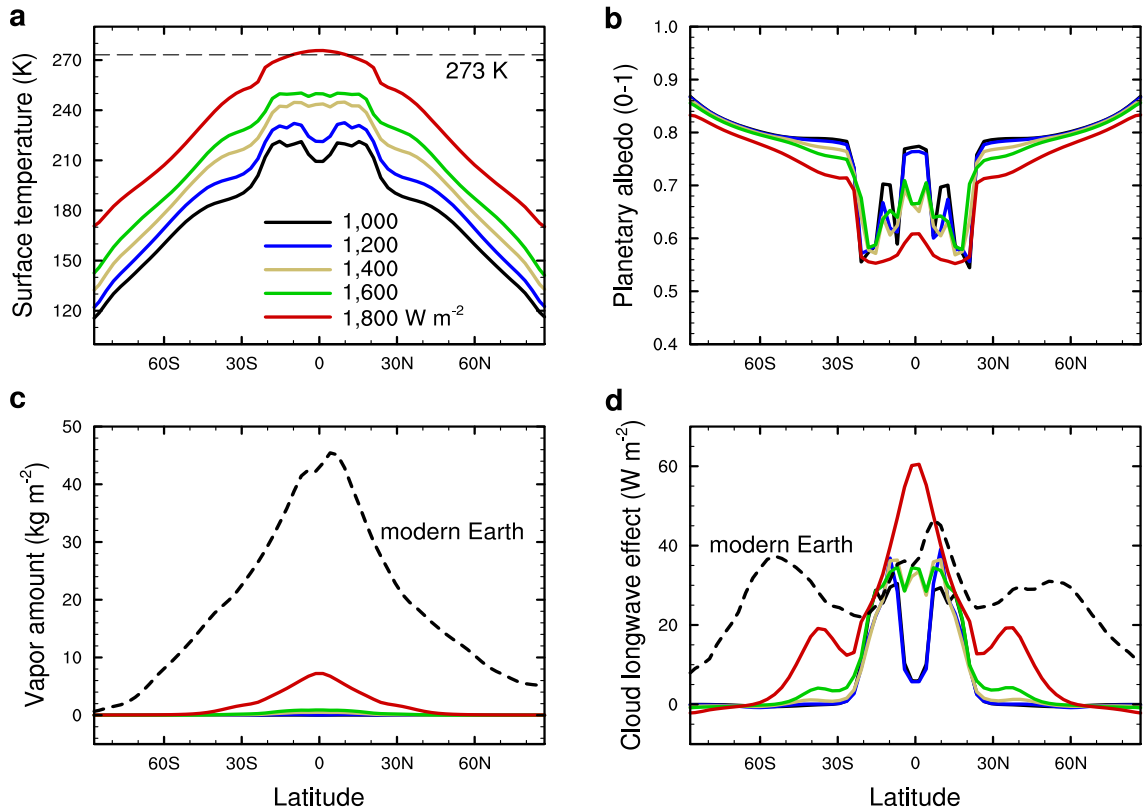

**Fig. 1.** Climates of a snowball planet around a G star. Annual- and zonal-mean surface temperature (**a**), planetary albedo (**b**), vertically integrated water vapor amount (**c**), and cloud longwave radiation effect at the top of the model (**d**). Stellar fluxes are 1,000, 1,200, 1,400, 1,600, and 1,800 W m$^{-2}$. The horizontal line in (**a**) is the melting point (273 K), and the dashed lines in (**c** & **d**) are the corresponding values for the modern Earth. In these cases, ice albedo is 0.5, snow albedo is 0.8, $CO_2$ amount is 300 ppmv, and obliquity is 0°.

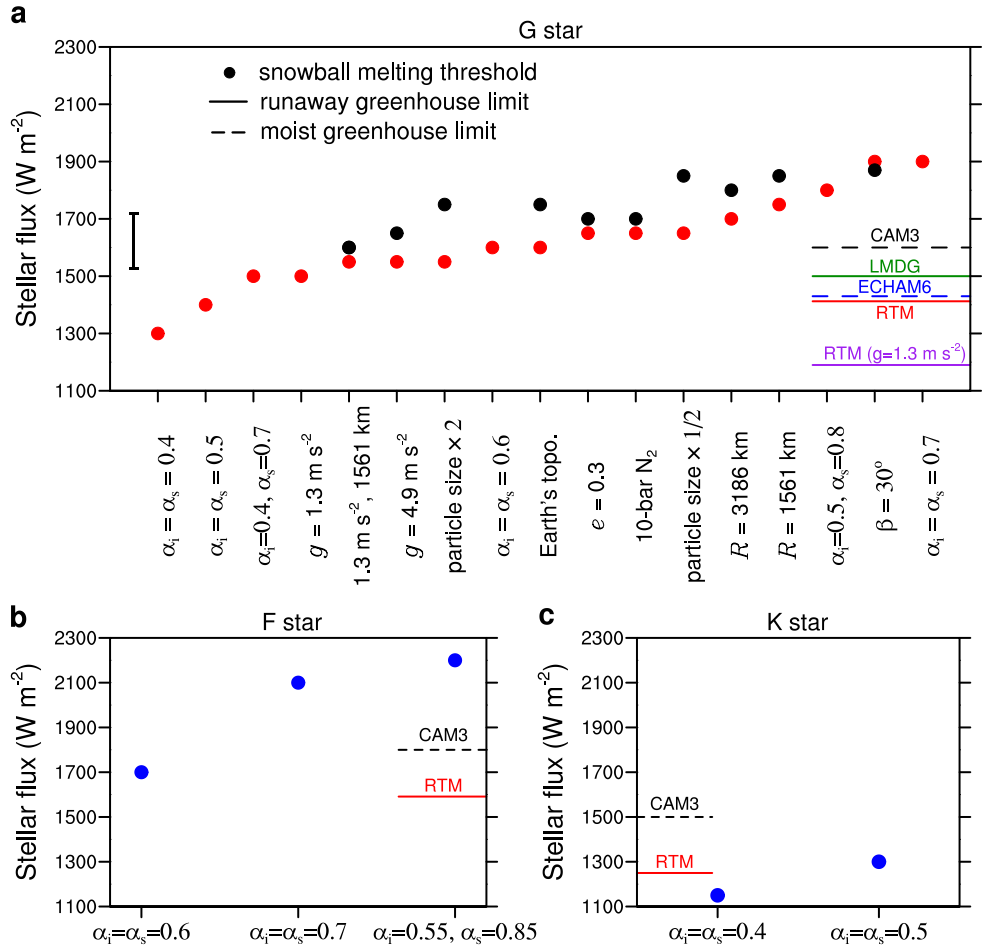

**Fig. 2.** Stellar flux thresholds for melting snowball planets and moons around G star (**a**), F star (**b**), and K star (**c**). Horizontal dashed and solid lines are the moist and runaway greenhouse limits, respectively, obtained by the 3D GCM CAM3[26], CAM4_Wolf[27], ECHAM6[28], LMDG[29], and 1D radiative transfer model (RTM)[30,31]. The vertical bar in (**a**) is the uncertainty range (170 W m$^{-2}$) in radiative transfer and clouds in estimating the stellar flux threshold (see Methods). Sensitivity parameters include ice albedo ($\alpha_i$), snow albedo ($\alpha_s$), gravity ($g$), radius ($R$), eccentricity ($e$), obliquity ($\beta$), ice cloud particle size, topographic height, and N$_2$ partial pressure. $\alpha_i = \alpha_s = 0.6$ (red dots), $\alpha_i = 0.5$ and $\alpha_s = 0.8$ (black dots), except if mentioned otherwise. For a lower gravity of 1.3 m s$^{-2}$, the runaway greenhouse limit is smaller, 1190 W m$^{-2}$.

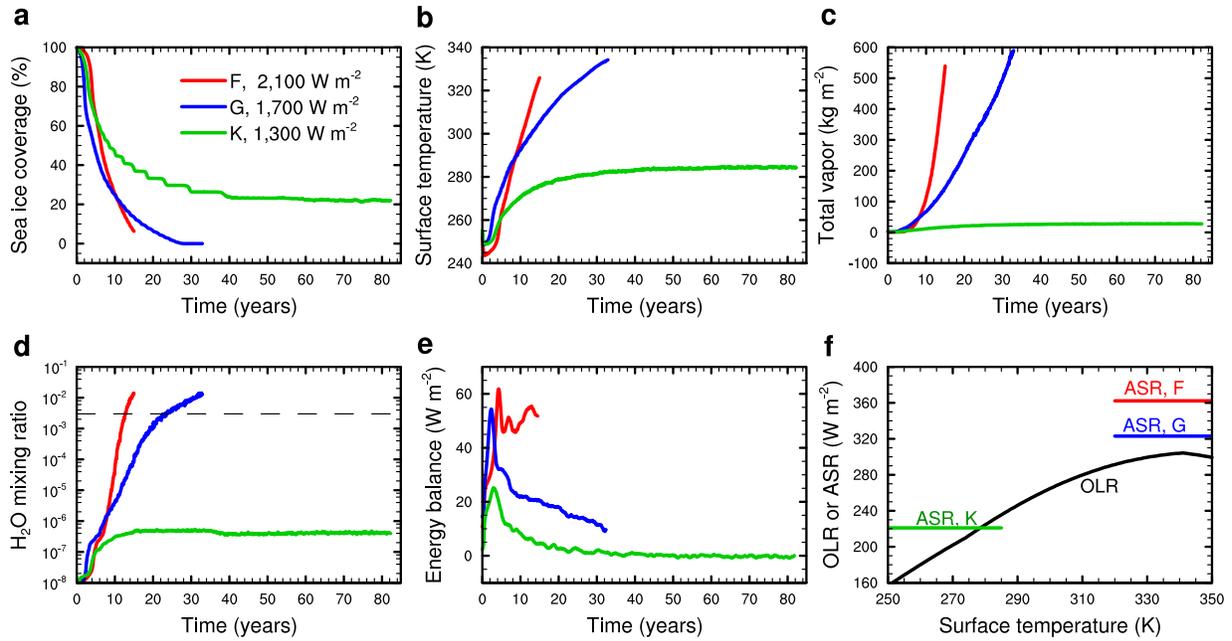

**Fig. 3.** Climate evolution of a snowball planet. Global-mean sea ice coverage (**a**), surface temperature (**b**), vertically integrated water vapor amount (**c**), water vapor volume mixing ratio in the upper atmosphere of the model (50 hPa, ~24 km in altitude when the surface temperature reaches 330 K, **d**), and energy balance at the top of the model (**e**, absorbed shortwave radiation (ASR) minus outgoing longwave radiation (OLR)), as a function of model integration time. **f**, OLR as a function of surface temperature (black line, see Methods) and ASR (horizontal lines) from the last converged solutions of the three experiments. Red line: 2,100 W m$^{-2}$ and F star; blue: 1,700 W m$^{-2}$ and G star; and green: 1,300 W m$^{-2}$ and K star. In (**d**), the horizontal dashed line is the critical moist greenhouse level, $3 \times 10^{-3}$.

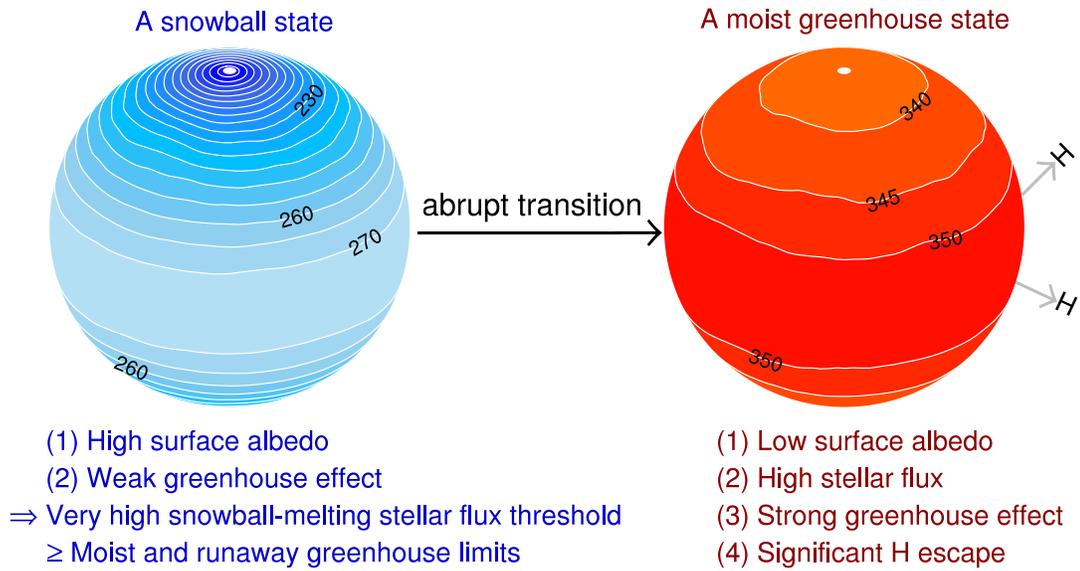

**Fig. 4.** Schematic illustration of the climate transition under stellar brightening and underlying physical mechanisms. The climate transition from a snowball state to a moist greenhouse state (or a runaway greenhouse state) is abrupt, dramatic, and irreversible. The contour lines are surface temperatures right before the snowball melting (left) and after the snowball melting (right), with a contour interval of 5 K.

**Acknowledgements:** We are grateful to D. S. Abbot, D. D.B. Koll, A. P. Showman & Z. Liu for their helpful discussions, and to J. Haqq-Misra for providing the OLR fitting coefficients. We thank Andrew Ingersoll and an anonymous reviewer for their helpful comments and suggestions. We thank the Editor Tamara Goldin for her help in improving the title and the first paragraph. J.Y. is supported by the National Science Foundation of China (NSFC) grants 41675071 and 41606060, Y.H. is supported by NSFC grants 41375072 and 41530423, and W.R.P. is supported by the Natural Sciences and Engineering Research Council of Canada Discovery Grant A9627. R.M.R acknowledges support by the Simons Foundation (SCOL 290357, Kaltenegger). The required computations were performed on the SciNet facility at the University of Toronto, which is a component of the Compute Canada HPC platform.

**Author Contributions** J.Y. led this work. J.Y., D.F. & Y.H. designed the study. J.Y. performed, analyzed and interpreted the simulations. D.F., R.M.R., W.R.P., Y.H. & Y.L. assisted with the analysis and interpretation of the data. W.R.P. provided computer code necessary for the simulations. All authors discussed the results and wrote the paper.

**Additional information**
Supplementary information is available in the online version of the paper. Reprints and permissions information is available online at www.nature.com/reprints. Correspondence and requests should be addressed to Y.J. (junyang@pku.edu.cn) and Y.H. (yyhu@pku.edu.cn).

**Competing financial interests**
The authors declare no competing financial interests.

**Methods**
**Climate model employed.** We employ the 3D Community Atmospheric Model version 3, CAM3[41]. The model solves the primitive equations for atmospheric motion on a rotating sphere and the parameterized equations for radiative transfer, convection, condensation, precipitation, clouds, and boundary turbulence[42]. The model has a spectral Eulerian core, a horizontal grid with a resolution of 2.8°, and 26 vertical levels from the surface to the middle stratosphere (3 hPa, i.e., ~46 km in altitude when the surface temperature is 330 K). A modified version of the model, in which the number of the vertical levels is increased to 34 and the pressure at the top of the model is extended to 0.2 hPa (i.e., ~69 km in altitude when the surface temperature is 330 K), has also been employed (see Fig. S3(c) below).

**Finding the snow-melting stellar flux threshold.** To find the stellar flux threshold for melting a snowball planet, stellar flux at the substellar point is specified as a series of values from 800 W m$^{-2}$ to a critical level in which the maximum annual-mean surface temperature reaches 273 K. The concentrations of atmospheric greenhouse gases ($CO_2$, $CH_4$, and $N_2O$) are fixed. An Earth-like atmosphere is employed, except if mentioned otherwise. Many parameters have also been examined, including ice and snow albedos, eccentricity, obliquity, gravity, radius, ice cloud particle size, surface topographic height (note surface albedo is set to being the same for different heights), and background gas concentration ($N_2$) (see Supplementary Table 1 online). Note that in a snowball state the spatial distribution of snow and bare ice depends on the strength of the seasonal cycle[43] and therefore on the planetary obliquity. Sensitivity tests show that our conclusion is robust in varying the obliquity (Fig. 2a). When the gravity or $N_2$ partial pressure is changed, the absolute masses (rather than mixing ratio) of $CO_2$, $CH_4$ and $N_2O$ are fixed. The snowball state is initialized with a surface temperature of 250 K everywhere, the time step in the model is 900 s, and each experiment is integrated for 20-40 model years.

Comparing with Shields et al.[9], the snowball-melting stellar flux threshold found in this study is higher, mainly because of two facts. First, the vertical resolution of the ice module used in Shields et al. is too low to resolve diurnal and seasonal cycles (see Figures 1-3 in ref. [44]). This induces unrealistic ice melting during midday but no nighttime refreezing, calling "a melt-ratchet effect"[16,44]. In our simulations, we set the surface to be glacial ice[45] (similar to that of Greenland in modern simulations) rather than sea ice[46]. The glacial ice is represented as ten vertical layers, and its resolution is much finer than that of sea ice, which is thereby able to avoid the melt-ratchet effect. Second, the ice and snow albedos used in Shields et al. are likely lower than should be in a snowball state. Shields et al. employed the sea ice and snow albedos on earth's polar regions, which are inappropriate for simulating the ice in the cold, dry snowball state--a fundamentally different climatic regime[47-49]. For example, Shields et al. used a sea ice albedo of 0.48 under the solar spectrum, which is lower than the suggested sea glacier albedo for snowball climate simulations, 0.55-0.65[47]. Both these two facts make Shields et al. underestimate the snowball-melting stellar flux threshold. In comparison, the ice and snow albedos employed in our simulations are more reasonable.

**Accuracy in low-temperature radiative transfer and snowball cloud simulations**. Comparing with line-by-line radiative transfer models, CAM3's radiative transfer scheme is accurate for surface temperatures being equal to or less than 310 K[50]. At 273 K, differences in longwave and shortwave radiation fluxes under clear-sky condition between CAM3 and the two line-by-line radiative transfer models are less than 5 W m$^{-2}$. The upper limit of $CO_2$ amount that CAM3 can simulate is ~100,000 ppmv[15]. At this $CO_2$ level, differences in clear-sky outgoing longwave radiation between CAM3 and Kasting's radiative transfer model[51] are within 2 W m$^{-2}$. We employ the solar spectrum and two stellar spectra corresponding to F-type (blackbody 7,200 K) and K-type (4,500 K) stars. Integrated broadband radiative flux difference[50] between a realistic spectrum and its corresponding blackbody spectrum is less than 7 W m$^{-2}$.

Through comparison with four state-of-the-art GCMs (CAM3, LMDz4, ECHAM6, and SP-CAM), a previous study[18] had established the fact that differences in net cloud radiative effect for Neoproterozoic snowball Earth simulations are within 10 W m$^{-2}$ and differences in tropical and annual mean surface temperature are within 5 K. Moreover, differences in the cloud radiative effect[19] between CAM3 and a cloud-resolving model (SAM) are within 2 W m$^{-2}$. These results imply that CAM3 is appropriate for simulating snowball clouds.

Combining the uncertainties in clear-sky and cloudy conditions, we state that the maximum uncertainty for simulating a snowball climate is within 17 W m$^{-2}$. This implies that the uncertainty range in estimating the stellar flux threshold for a snowball melting is ~170 W m$^{-2}$ (i.e., $4 \times 17/(1-\alpha_p)$, where $\alpha_p$ is planetary albedo (such as 0.6) and the factor of 4 is the ratio of the planet's surface area to its cross-sectional area). This uncertainty is shown as a vertical bar in Fig. 2a of the main text.

**Moist and runaway greenhouse limits.** Both 1D radiative-transfer models and 3D GCMs have previously been employed to estimate the moist and runaway greenhouse limits. For a pure water vapor atmosphere and a surface albedo of 0.2, the runaway greenhouse limit under the solar spectrum is ~1,340 W m$^{-2}$ in SMART[31] and ~1,420 W m$^{-2}$ in the 1D radiative-transfer model LMDG[29] and Kasting's radiative transfer model[30]. For an Earth-like atmosphere with saturated water vapor, the runaway greenhouse limit is ~1,430 W m$^{-2}$ with an uncertainty range[50] of 10%. For a F-star spectrum, the moist and runaway greenhouse limits are higher, and for a K-star spectrum, the limits are lower[8,26,30].

The runaway greenhouse limit in the 3D GCM LMDG[29], ~1,500 W m$^{-2}$, is higher that in the 1D radiative-transfer model LMDG, mainly due to the effects of atmospheric sub-saturation. The moist greenhouse limit is ~1,430 W m$^{-2}$ in ECHAM6[28] and ~1,600 W m$^{-2}$ in the GCMs CAM3[26] and CAM4_Wolf[27]. Divergences between the 3D GCMs are due to the differences in cloud parameterization[28] and shortwave absorption by water vapor[50].

A planet with weaker gravity has a lower runaway greenhouse limit than one with stronger gravity. Based on this, Pierrehumbert speculated that moons in Earth-like orbits are in either a runaway greenhouse state or a snowball state[12]. Note that moons as well as small planets are able to maintain long-lived and substantial atmospheres as long as their masses are greater than ~0.1-0.2 Earth's mass[52-55].

**Climate transition simulations.** In examining the post-snowball climate, CAM3 is coupled to a thermodynamic sea ice module[46] and to a 50-m slab ocean module[41]. Sea ice can grow and melt but glacier ice cannot in the model. Stellar flux is fixed to the typical stellar flux threshold for a snowball melting. No ocean heat transport has been included. The effects of stellar spectrum, planetary mass, $CO_2$ concentration and cloud particle size on the results are investigated; the post-snowball Earth climate that occurred in ~600-800 million years ago is also examined (see Supplementary Table 2 & Fig. S8). The experiments have an initial surface temperature of 250 K everywhere (a snowball state), and the time step is 60 s in the simulations using the standard version of the model and 10 s in the simulations using the modified version of the model that has a top pressure of 0.2 hPa. Each case is integrated for 80-120 model years.

**Uncertainty in high-temperature radiative transfer**. Comparing with line-by-line radiative transfer models, CAM3 underestimates the greenhouse effect and shortwave absorption of water vapor at high temperatures[50], and the model crashes when the maximum surface temperature is higher than 340 K due to numerical instability and radiative transfer limitation[56]. For example, at 320 K, the greenhouse effect and shortwave absorption in CAM3 is 8-11 W m$^{-2}$ weaker than that in the two line-by-line models SMART and LBLRTM[50]. This implies that the surface temperature in the post-snowball state should be greater and stratospheric vapor concentration should be higher than those shown in Figs. 3 & S3. Future simulations including high-temperature radiative transfer and $H_2O$ photolysis are required to more precisely investigate the post-snowball climates.

**Outgoing longwave radiation (OLR) curves and absorbed stellar radiation (ASR) values.** At low temperatures, OLR is a nearly linear function of surface temperature, and at high temperatures, OLR asymptotes to a limiting, maximum value when the atmosphere becomes optically thick in all infrared wavelengths. Data for the OLR curves in Figs. 3f and S9 are from Haqq-Misra *et al.*[57], and we have further included the effects of atmospheric sub-saturation[29,31] and clouds. This limiting OLR is approximately equal to the blackbody radiation corresponding to the temperature at the pressure level of one optical depth calculated down from the top of the atmosphere[12]. For pure water vapor, the limiting OLR[31] is ~282 W m$^{-2}$; for an atmosphere of 1-bar $N_2$ and saturated water vapor[29,31], it is ~292 W m$^{-2}$; the realistic atmosphere is sub-saturated and the limiting OLR would be higher, such as ~340 W m$^{-2}$ if assuming relative humidity is 45%, close to the equivalent value in 3D GCMs[26,27]; clouds can decrease the limiting OLR by ~30 W m$^{-2}$ (obtained in the CAM3 simulations of this work) through trapping infrared radiation from the surface. The limiting OLR for an atmosphere with 1-bar $N_2$, sub-saturated water vapor and clouds, therefore, is ~310 W m$^{-2}$.

The horizontal ASR lines in Fig. 3f of the main text are calculated based on the stellar flux thresholds (2,100, 1,700, and 1,300 W m$^{-2}$ for F, G, and K stars, respectively) and planetary albedos (0.32, 0.24, and 0.34) from the last converged solution of CAM3 post-snowball experiments shown in Fig. 3a-e of the main text. The existence of a limiting OLR and ASR being larger than the limiting OLR are two necessary conditions for the onset of an unstable runaway greenhouse state[8,12].

**Effects of atmospheric $CO_2$ on the stellar flux threshold and the post-snowball climate:** How much atmospheric $CO_2$ could accumulate in a snowball state is determined by the balance between silicate (and sea floor) weathering and volcanic outgassing. In order to know the effect of $CO_2$ on the results, we perform five groups of experiments (Fig. S6). When the surface albedo is set to 0.6 and $CO_2$ concentration is set to < 3,000 ppmv, the snowball-melting stellar flux threshold is still higher than the runaway greenhouse limit[29]. This implies that the post-snowball planet will enter a runaway greenhouse state. If $CO_2$ concentration is > 3,000 ppmv, the post-snowball climate will be hot, but do not enter a runaway greenhouse state because the stellar flux threshold is lower than the runaway greenhouse limit. When the surface albedo is set to 0.8 for snow and 0.6 for sea ice and $CO_2$ concentration is < 30,000 ppmv, a runaway greenhouse state after the snowball melting will also arise. But, if the $CO_2$ concentration is > 30,000 ppmv, it will not.

An example of the post-snowball climate with a high concentration of $CO_2$, 100,000 ppmv, is shown in Fig. S7. In this case, the snowball-melting stellar flux threshold is only ~1,360 W m$^{-2}$, the corresponding post-snowball global-mean surface temperature is 325 K, and stratospheric $H_2O$ amount is somewhat lower than that required for entering a moist greenhouse state. However, the surface temperature and stratospheric $H_2O$ amount are expected to decrease after the snowball deglaciation, because enhanced silicate weathering[58] following the melting would lead to the removal of atmospheric $CO_2$. Therefore, icy planets near the outer edge of the habitable zone that have an active carbon-silicate cycle may possess repeated climate cycles between the snowball state and the hot state[7,57]. This is different from the process of stellar brightening, in which the planet will enter a moist or runaway greenhouse state accompanying by significant water loss from the ocean(s) and it has no chance to return back to a normal climate state.

**A snowball Earth:** The climate evolution associated with a snowball Earth 600-800 million years ago is summarized in Fig. S8 (refs. 4, 16, 17, 18, 59-66). As shown in the figure, the climate transition from the snowball Earth state to the post-snowball state is abrupt and dramatic, but the post-snowball state is in neither a moist greenhouse state nor a runaway greenhouse state. The reasons are: (1) The solar constant, 1286 W m$^{-2}$ (~6% fainter), is much lower than the runaway greenhouse limit[29] of ~1500 W m$^{-2}$. Here we assume that atmospheric $CO_2$ concentration does not significantly influence the onset of a runaway greenhouse state[8]; this is because near the runaway greenhouse state $H_2O$ becomes a major component of the atmosphere and dominates its infrared opacity. (2) After the snowball Earth melts, the surface is hot but stratospheric $H_2O$ mixing ratio is not higher than the moist greenhouse limit of 3 x 10$^{-3}$. This is due to the fact that $CO_2$ acts to warm the surface and troposphere but to cool the stratosphere and thereby decrease $H_2O$ mixing ratio there[40]. (3) Silicate weathering following the melting would enhance due to increased precipitation and runoff[58], which would lead to the removal of $CO_2$. The surface temperature and stratospheric $H_2O$ amount are therefore expected to decrease after the hot period. The time scale of silicate weathering for dropping down the $CO_2$ concentration of ~100,000 ppmv to a normal level is ~10$^6$ years[58]. Even if assuming the stratospheric $H_2O$ mixing ratio during the entire hot period were 3 x 10$^{-3}$, the loss of ocean water would be only 0.6 m, very tiny.

The climate evolution associated with early Mars and Venus can be inferred from the results for the snowball Earth. The solar constant on Mars is 591 W m$^{-2}$, and in the history it was lower. So that, in order to melt a snowball Mars if it was, $CO_2$ concentration should be much higher than the $CO_2$ level for the snowball Earth melting. Due to the fact that $CO_2$ acts to cool the stratosphere, a post-snowball Mars would not enter a moist greenhouse state. A snowball Mars would also not transition to a runaway greenhouse state because its solar flux was much lower than the runaway greenhouse limit[12] of ~1250 W m$^{-2}$ under the Mars gravity of 3.71 m s$^{-2}$. Moreover, it is doubtful that Mars was ever in a snowball state. The current estimates for its initial global water inventory, < 165 m global water equivalent[67], suggest that the planet was never as wet as Earth. Climate simulations shown only about 30% of its surface was glaciated[68]. The solar constant on Venus is ~2630 W m$^{-2}$, and in four billion years ago it was ~1970 W m$^{-2}$. Both of these two insolation levels are higher than the stellar flux threshold for a snowball melting (Fig. 2(a)). This implies that it is unlikely that Venus had been in a snowball state.

**Energy balance climate model:** The climate transition and the effect of atmospheric greenhouse gas concentration on the transition can be further understood using a 0D energy balance climate model. The key processes of the model are ice-albedo feedback and water vapor feedback. Due to these two processes, there are multiple equilibrium states for a climate system (ref. chapter 3 of reference 12).

The balance between absorbed stellar radiation (ASR) and outgoing longwave radiation (OLR) determines the equilibrium surface temperature, as shown in Fig. S9. In the figure, the ASR curves are calculated as $S_0/4 \times (1-\alpha_p)$, where the albedo ($\alpha_p$) is 0.25 when the surface temperature is higher than 300 K and 0.60 when the surface temperature is lower than 260 K. Between the two temperatures, a quadratic interpolation[12] for the albedo is employed. For a relatively low stellar flux ($S_0$ = 1,400 W m$^{-2}$, the blue line in panel a), there are three solutions: a stable snowball state, an unstable partially ice-covered state, and a stable ice-free state. For a relatively high stellar flux ($S_0$ = 1,700 W m$^{-2}$, the red line in panel a), there are two solutions: a stable snowball state and an unstable partially ice-covered state. There is no equilibrium ice-free state for the case of $S_0$ = 1,700 W m$^{-2}$; as long as all of the surface ice melts, ASR would be higher than the allowed maximum OLR and the system would enter a runaway greenhouse state, which corresponds to the dramatic climate transition obtained in the 3D model CAM3.

If $CO_2$ concentration is high, the OLR curve shifts downward (Fig. S9(b)), due to the $CO_2$ greenhouse effect. For the case of $S_0$ = 1,400 W m$^{-2}$, there are still three solutions but the two stable solutions (snowball and ice-free) are warmer than those in the case of a low $CO_2$ concentration. This is due to the $CO_2$ greenhouse effect. For the case of $S_0$ = 1,700 W m$^{-2}$, there is no equilibrium solution because the high $CO_2$ greenhouse effect prevents the system from staying in a snowball state, meanwhile the high stellar flux pushes the system into a runaway greenhouse state.

**K and M stars**: The abrupt climate transition does not apply to K- and M-star systems, due to their redder spectra and therefore lower ice/snow albedos and smaller snowball-melting stellar flux thresholds. Moreover, due to strong tidal forces, planets in the habitable zone of late K- and M-star systems are likely in spin-orbit resonance states[69], which has a profound impact on planetary climate[32]. A snowball planet in 1:1 resonance state (or called synchronous rotation) would require a much lower stellar flux to trigger the melting because all of the stellar energy is deposited on the permanent day side, compared with other resonance states.

**Data Availability**: The data that support the findings of this study are available in the PANGAEA repository, https://doi.pangaea.de/10.1594/PANGAEA.876224.

**Code Availability**: The code used to generate the data for this study is available in the PANGAEA repository, https://doi.pangaea.de/10.1594/PANGAEA.876224.

**References Cited in Methods**

**Supplementary Information for**
**Abrupt Climate Transition of Icy Worlds from Snowball to Moist or Runaway Greenhouse**

by Jun Yang, Feng Ding, Ramses M. Ramirez, W. R. Peltier, Yongyun Hu, and Yonggang Liu

**This PDF file includes:**

1. Supplementary Tables S1 & S2
2. Supplementary Figures S1-S9

**Table S1: Summary of the snowball climate simulations using the model CAM3 coupled to a glacier ice module**

| Groups[1] | Runs | Description |
|---|---|---|
| Control | 5 | The surface is set to have a sea ice albedo of 0.5 and a snow albedo of 0.8. The partial pressure of $N_2$ is ~1 bar; $CO_2$ concentration is 300 ppmv; there are no $O_2$, $O_3$, CFCs or aerosols; and surface topographic height is zero. Gravity is 9.8 m s$^{-2}$; planetary radius is 6,372 km; rotation period is 1 day; orbital period is 365 days; and both obliquity and eccentricity are zero. Stellar fluxes[2] are 1,000, 1,200, 1,400, 1,600 and 1,800 W m$^{-2}$, under the G-star spectrum. |
| Stellar spectrum | 20 | Two stellar spectra corresponding to F-type (blackbody 7,200 K) and K-type (4,500 K) stars are employed. For F star, the surface albedo is set to have a uniform value of 0.6 or 0.7, or a sea ice albedo 0.55 and a snow albedo of 0.85. For K star, the surface albedo is set to have a uniform value of 0.4 or 0.5. |
| Surface albedo | 24 | The surface has a constant albedo of 0.4, 0.5, 0.6 or 0.7, or has a sea ice albedo of 0.4 and a snow albedo of 0.7. |
| Planetary mass | 16 | The gravity is decreased to 4.9 or 1.3 m s$^{-2}$, and/or the planetary radius is decreased to 3,186 or 1,561 km, in order to examine small-size planets and moons. |
| $CO_2$ concentration | 20 | The $CO_2$ concentrations are set to 3, 30, 3,000, 30,000 and 100,000 ppmv, respectively, in order to examine the sensitivity of the results to $CO_2$ amount. |
| Orbit geometry | 8 | The eccentricity is increased from 0 to 0.3, or the obliquity is increased from 0° to 30°. |
| Atmospheric mass | 4 | The partial pressure of $N_2$ is increased from 1 bar to 10 bars[3], in order to examine the effect of background gas concentration. |
| Cloud particle size | 8 | The cloud ice particle sizes are set to 2 times or half of those in the control case. Earth GCMs have large uncertainties in cloud simulations. |
| Surface topography | 4 | The topographic height is set to the values of the present-day Earth. |

[1]The results are shown in Figs. 1, 2, S1, S2, and S3.

[2]For the control group as well as other groups, each case is run for four or five different stellar fluxes in order to find the snowball-melting stellar flux threshold.

[3]If the background gas concentration is decreased, temperature at the tropopause will be higher and thereby the $H_2O$ mixing ratio at and above the tropopause will be larger, promoting the onset of a moist greenhouse state (refs. 72 & 73).

**Table S2: Summary of the post-snowball climate simulations using the model CAM3 coupled to a 50-m slab ocean module**

| Group[1] | Cases | Description |
|---|---|---|
| Control | 1 | The partial pressure of $N_2$ is ~1 bar; $CO_2$ concentration is 300 ppmv; and there are no $O_2$, $O_3$, CFCs or aerosols. Gravity is 9.8 m s$^{-2}$; planetary radius is 6372 km; rotation period is 1 day; orbital period is 365 days; and both obliquity and eccentricity are zero. The stellar flux is 1,700 W m$^{-2}$, under the G-star spectrum. |
| Stellar spectrum | 2 | Two stellar spectra corresponding to F-type (blackbody 7,200 K) and K-type (4,500 K) stars are employed. For F star, the stellar flux is 2,100 W m$^{-2}$. For K star, it is 1,300 W m$^{-2}$. |
| Planetary mass | 5 | The gravity is decreased to 4.9 or 1.3 m s$^{-2}$, and/or the planetary radius is decreased to 3,186 or 1,561 km, in order to examine small-size planets and moons. |
| $CO_2$ concentration | 1 | The $CO_2$ concentration is increased from 300 to 100,000 ppmv, and the snowball-melting stellar flux threshold is 1,360 W m$^{-2}$. |
| Cloud particle size | 2 | The cloud liquid and ice particle sizes are set to 2 times or half of those in the control case. |
| Snowball Earth | 4 | The $CO_2$ concentration is 20,000, 40,000, 100,000 or 200,000 ppmv, and the stellar flux is 1,286 W m$^{-2}$ (i.e., 94% of the present level). The surface continental configuration is that for ~630 million years ago (ref. 16). The planetary obliquity is 23.5º. |
| Model top pressure | 3 | Same as the control case, but the pressure at the top of the model is extended from ~3 to ~0.2 hPa and the number of model vertical levels is increased from 26 to 34. |

[1]The results are shown in Figs. 3, S4, S5, S6, S7, and S8.

**Supplementary Figures S1-S9:**

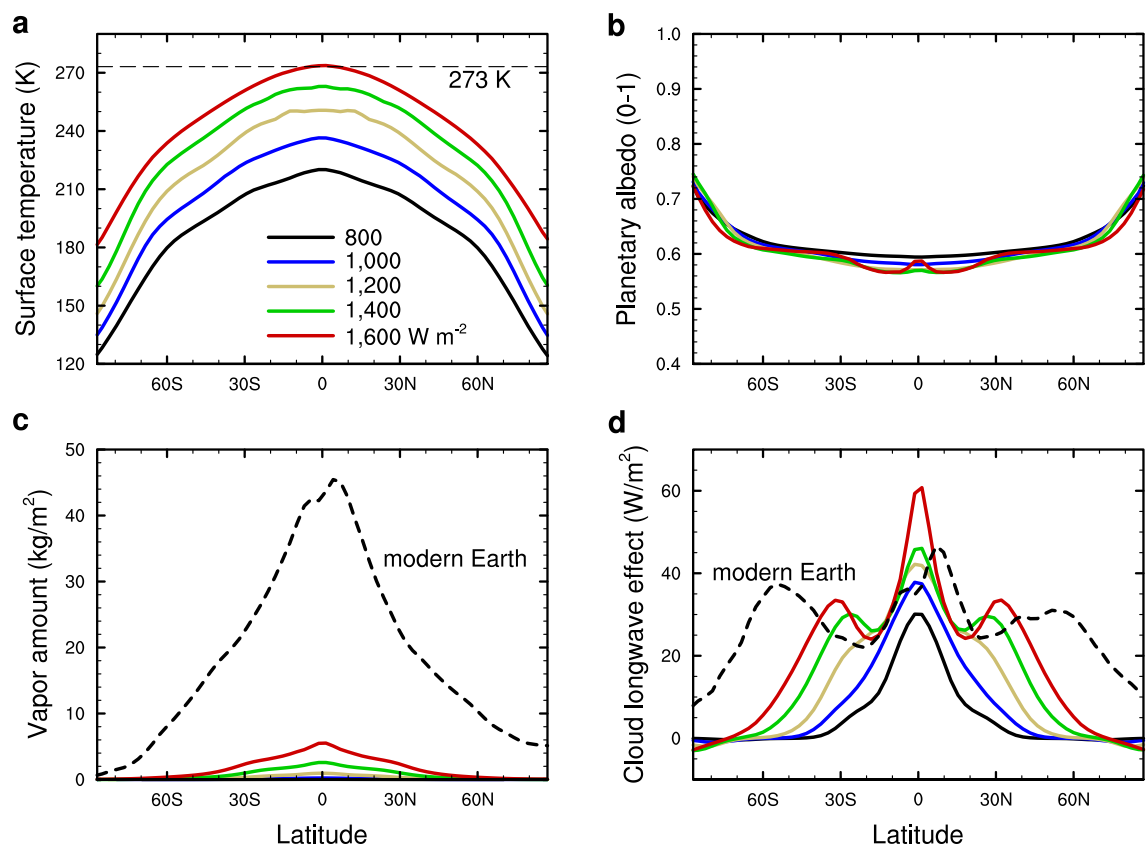

Fig. S1. Climates of a snowball planet around a G star. The surface has a uniform albedo of 0.6 and $CO_2$ concentration is 300 ppmv. Annual- and zonal-mean surface temperature (**a**), planetary albedo (**b**), vertically integrated water vapor amount (**c**), and cloud longwave effect at the top of the model (**d**), for different stellar fluxes from 800 to 1,600 W m$^{-2}$. The horizontal line in (**a**) is the melting point (273 K), and the dashed lines in (**c** & **d**) are the corresponding values of the modern Earth. The snowball-melting stellar flux threshold is 1,600 W m$^{-2}$ in this group of experiments.

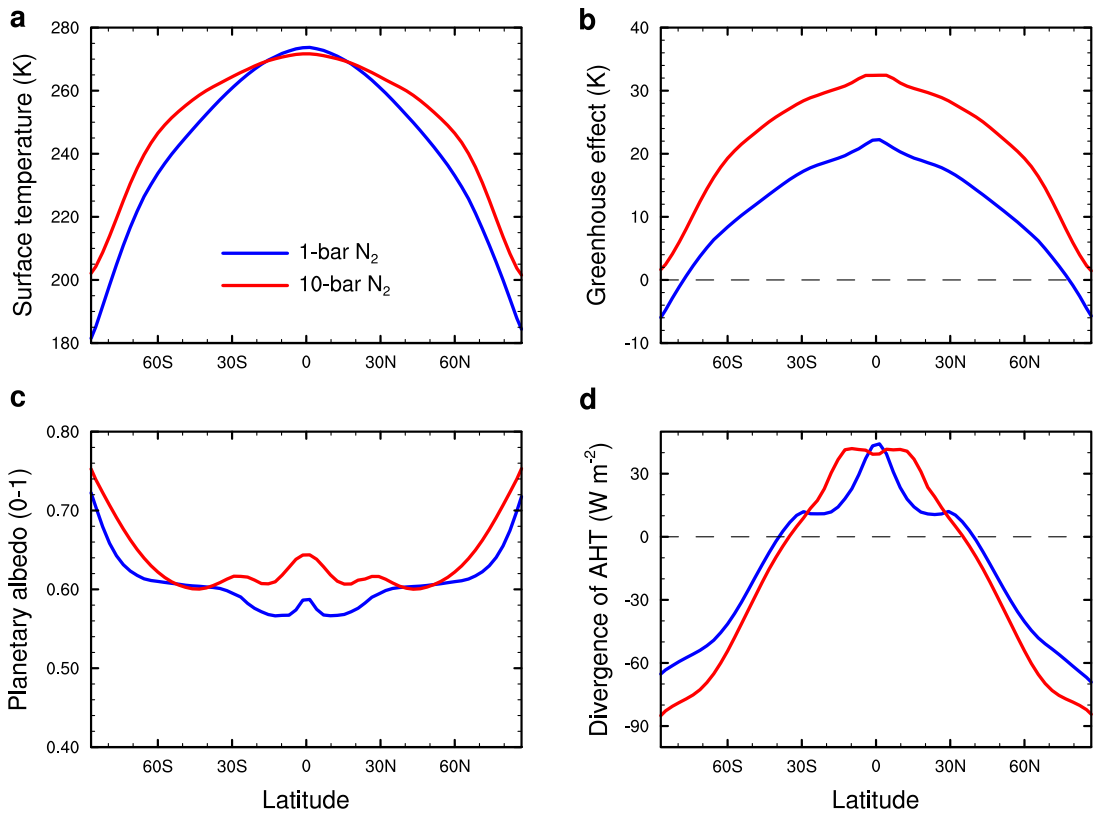

Fig. S2. Climates of a snowball planet under two $N_2$ partial pressures, 1 bar (blue line) and 10 bars (red). Annual- and zonal-mean surface temperature (**a**), greenhouse effect (**b**), planetary albedo (**c**), and divergence of meridional atmospheric heat transport (AHT, **d**). Increasing $N_2$ partial pressure raises the global-mean surface temperature through pressure broadening but has a small effect on the deep-tropical surface temperatures, primarily due to the increased planetary albedo and the enhanced AHT from the tropics to the middle and high latitudes. In these simulations, the stellar flux is 1,600 W m$^{-2}$, stellar spectrum is the Sun's, surface albedo is 0.6 (uniform), planetary obliquity is 0°, and $CO_2$ column content is 4.5 kg m$^{-2}$ (i.e., 300 ppmv under 1-bar $N_2$).

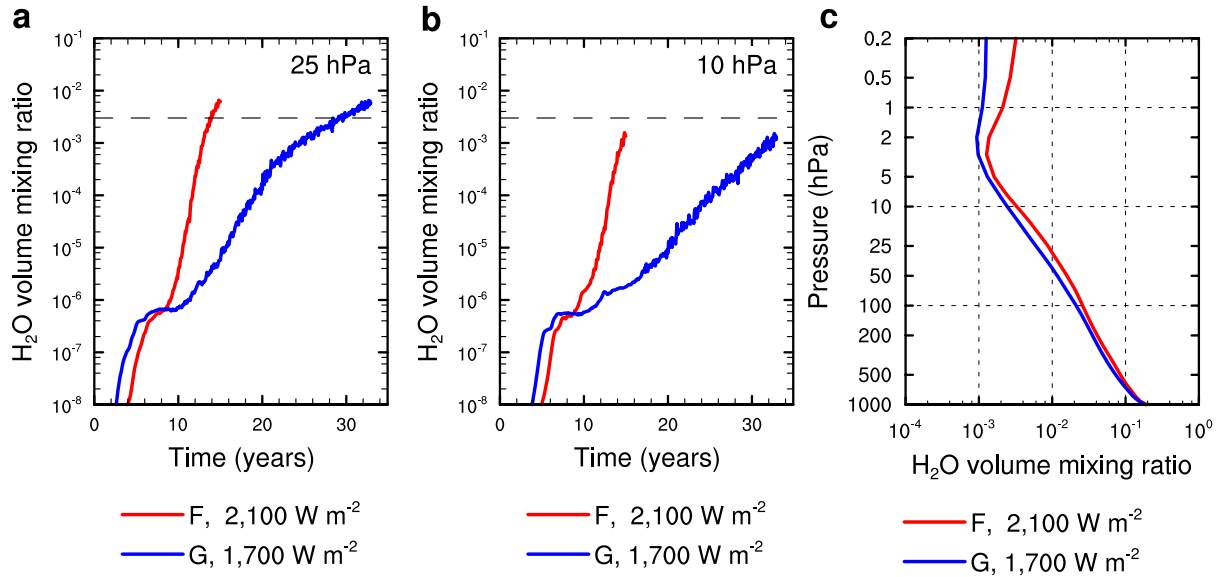

Fig. S3. Global-mean water vapor volume mixing ratio. (**a & b**) Water vapor volume mixing ratio as a function of model integration time in the standard version of the model, same as Fig. 3e in the main text, but for 25 hPa (~29 km in altitude when the surface temperature reaches 330 K, **a**) and 10 hPa (~36 km, **b**). In the standard version of the model, the pressure at the top of the model is 3 hPa. **c**, water vapor volume mixing ratio as a function of pressure in the last converged solution of the simulations using the modified version of the model, which has a pressure of 0.2 hPa (~69 km) at the top of the model. Red line: a F-star spectrum and a stellar flux of 2,100 W m$^{-2}$, and blue line: G-star and 1,700 W m$^{-2}$. In all of the three panels, the last converged water vapor volume mixing ratios in the upper atmosphere of the model are close to or higher than the critical moist greenhouse level of $3 \times 10^{-3}$ (the horizontal dashed line in **a & b**).

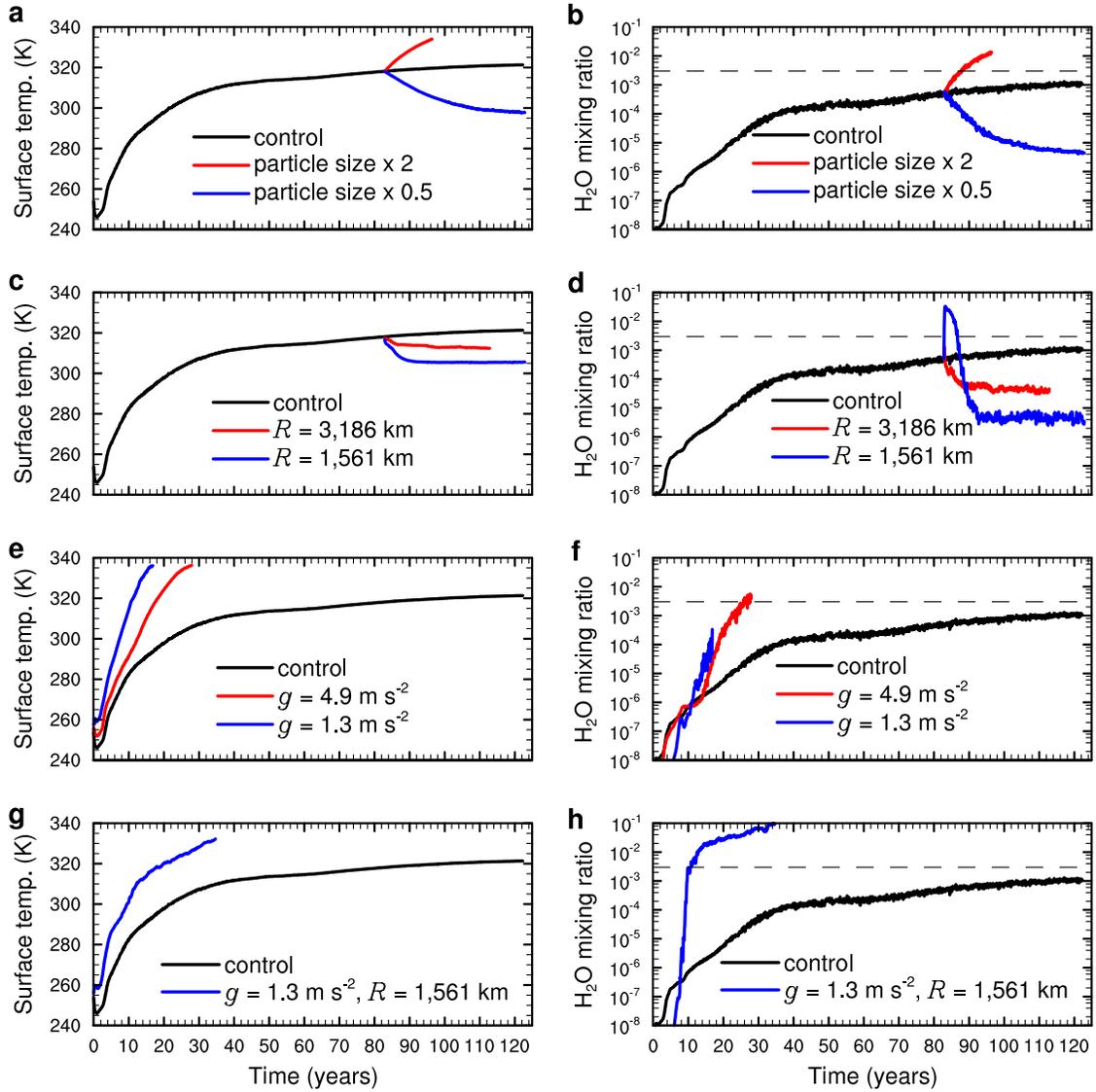

Fig. S4. Climate evolution of snowball planets. Left panels: Global-mean surface temperature, and right panels: Water vapor volume mixing ratio at 50 hPa for $g$ = 9.8 m s$^{-2}$ (25 hPa for $g$ = 4.9 m s$^{-2}$, or 6.6 hPa for $g$ = 1.3 m s$^{-2}$), in four groups of experiments. (**a & b**) varying liquid and ice cloud particle sizes; (**c & d**) varying planetary radius ($R$) only; (**e & f**) varying planetary gravity ($g$) only; and (**g & h**) varying both $R$ and $g$. In the control experiment, $R$ = 6,372 km, and $g$ = 9.8 m s$^{-2}$. The stellar flux is 1,600 W m$^{-2}$, N$_2$ partial pressure is 1 bar, the absolute mass of CO$_2$ is the same, and stellar spectrum is the Sun's, in all of the simulations. In **a-d**, the sensitivity experiments are restarted from the 83$^{rd}$ year of the control experiment. The horizontal lines in **b**, **d**, **f** & **h** are the critical moist greenhouse level, 3 × 10$^{-3}$.

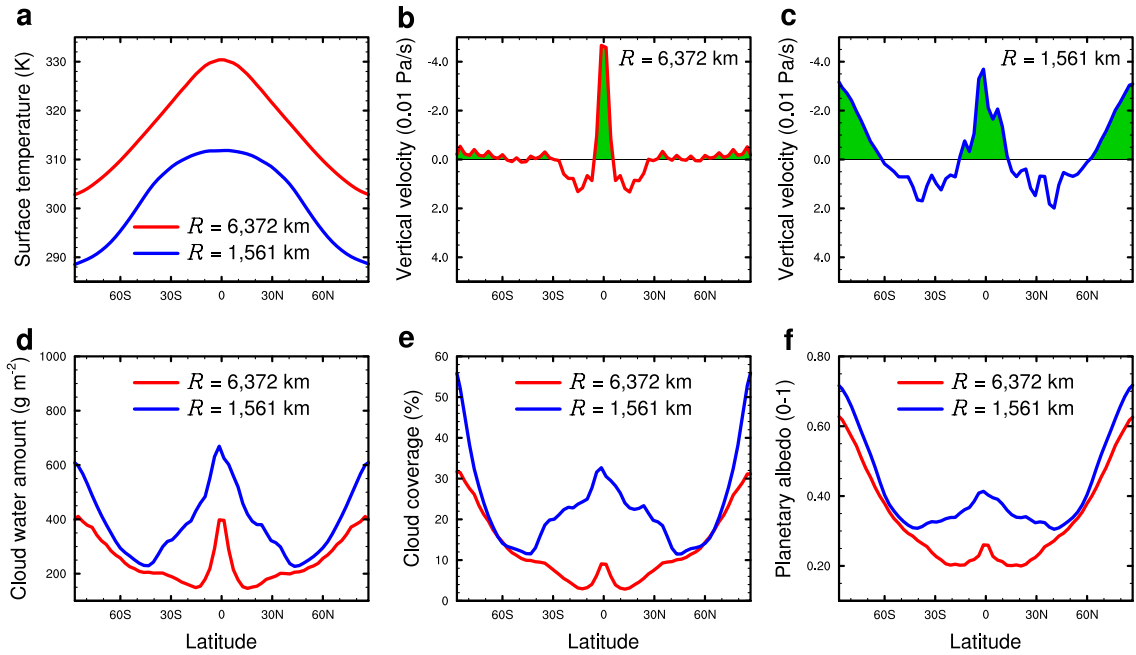

Fig. S5. Effect of planetary radius ($R$) on the post-snowball climate in a G-star system. Annual- and zonal-mean surface temperature (**a**), pressure-weight vertical average of vertical velocity (negative values denote upwelling, between 300 and 1,000 hPa) for $R$ = 6,372 km (**b**) and for $R$ = 1,561 km (Europa's value, **c**), vertically integrated cloud (liquid and ice) water amount (**d**), cloud coverage between 400 and 700 hPa (middle troposphere, **e**), and planetary albedo (**f**). In **b** & **c**, the region of upwelling motion is color-shaded. Stellar flux is 1,600 W m$^{-2}$, $g$ is 9.8 m s$^{-2}$, and the only difference between the two experiments is $R$. When $R$ is reduced, the width of the Hadley cells in degree (º) of latitude increases[70,71] and the width of tropical upwelling motion region increases, so that cloud water amount, cloud coverage and cloud optical thickness increase, and thereby planetary albedo increases and the surface becomes cooler.

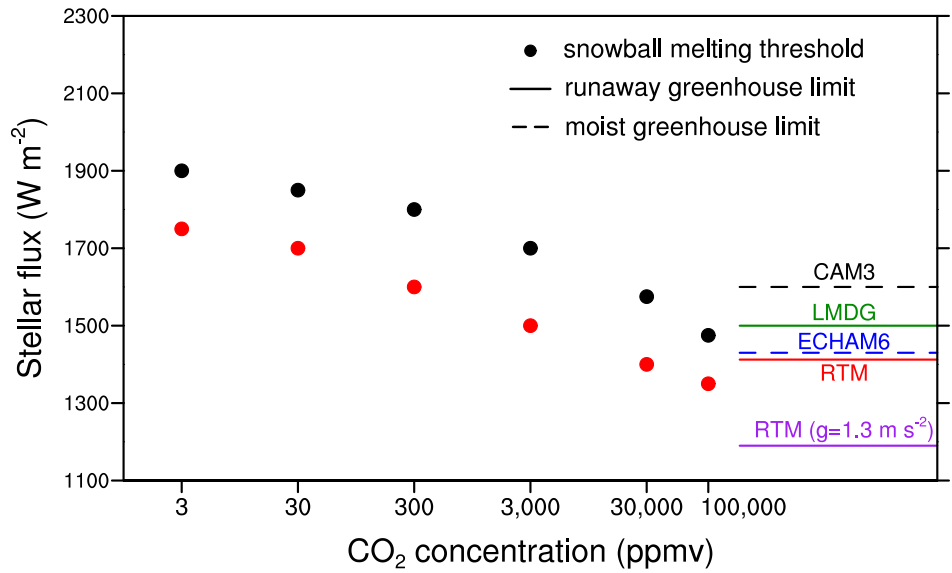

Fig. S6. The snowball-melting stellar flux threshold as a function of $CO_2$ concentration. The surface is set to have a constant surface albedo of 0.6 (red dots) or to have a snow albedo of 0.8 and a sea ice albedo of 0.5 (black dots). Horizontal dashed and solid lines are the moist and runaway greenhouse limits, respectively, obtained by various models (same as Fig. 2a).

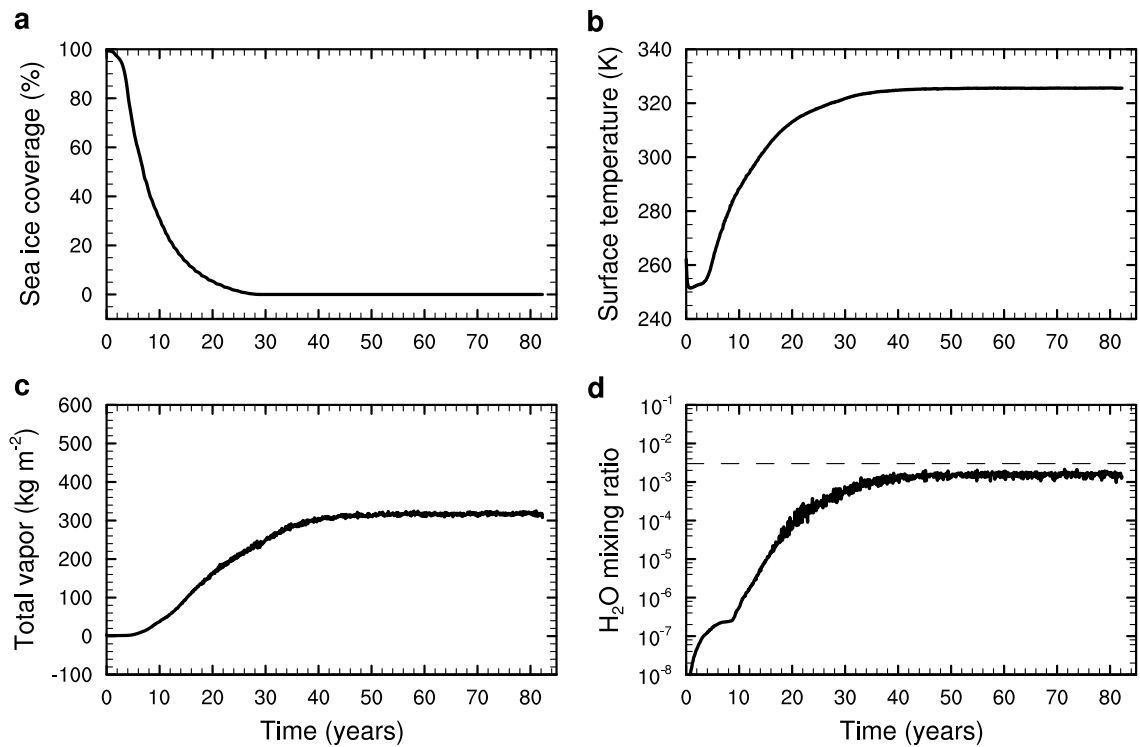

Fig. S7. Climate evolution of a snowball planet with a high concentration of $CO_2$, 100,000 ppmv. Global-mean surface ice coverage (**a**), surface temperature (**b**), vertically integrated water vapor amount (**c**), and water vapor volume mixing ratio at 50 hPa (~24 km in altitude when the surface temperature reaches 330 K, **d**). The stellar flux is 1,360 W m$^{-2}$, stellar spectrum is the Sun's, and planetary obliquity is 23° in this simulation.

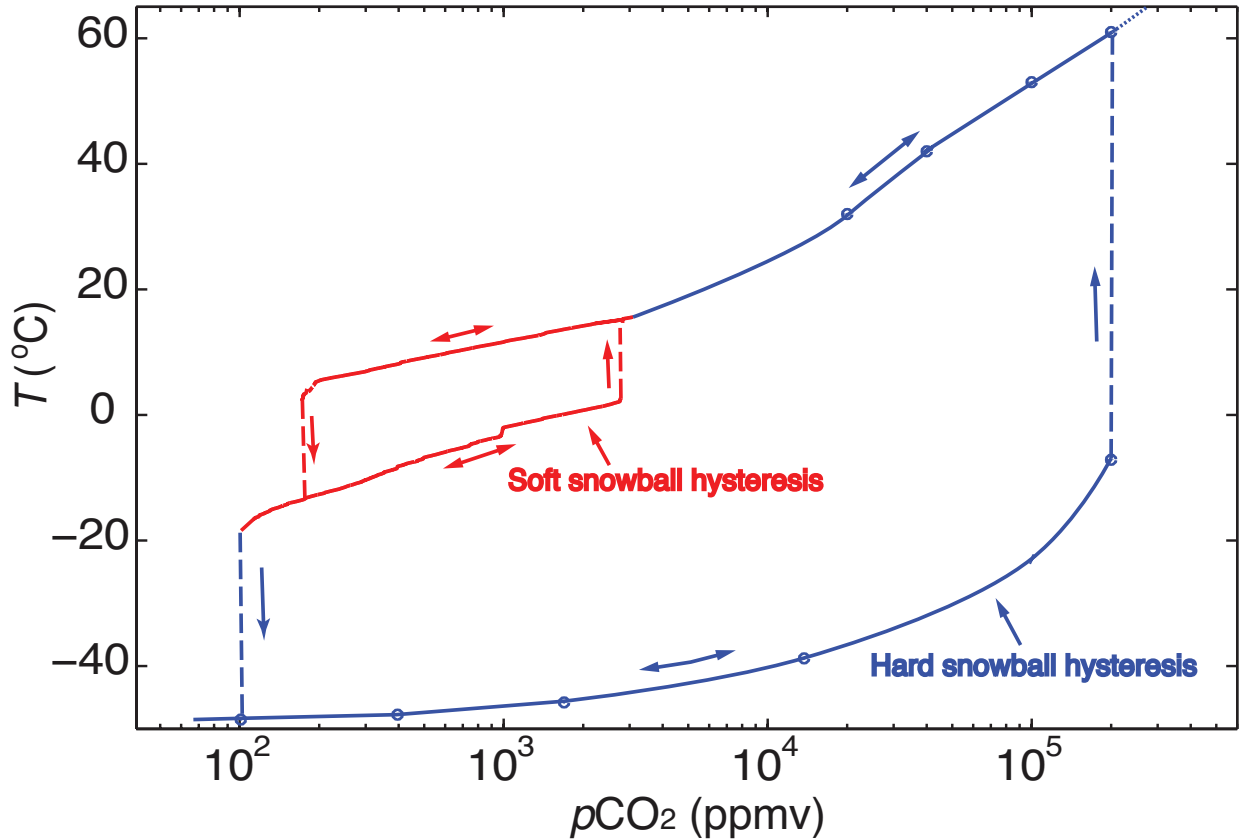

Fig. S8. Hysteresis loop of the global-mean surface temperature as a function of $CO_2$ concentration for the hard snowball Earth hypothesis (blue lines) and the soft snowball Earth hypothesis (red lines). In the hard snowball state, continental ice sheets reach the deep tropics and the oceans are completely covered by thick ice (refs. 4, 16). The hard snowball hysteresis was simulated using CAM3 for the climates in and after the hard snowball (refs. 17-18, and this study) and using the coupled atmosphere-ocean version of CAM3--CCSM3 for the initiation of a snowball Earth (refs. 59-63). In the soft snowball state, ice-free oceans in the tropics coexist with low-latitude continental glaciers and mid- and high-latitude sea ice, which was simulated with a coupled ice sheet--energy balance climate model (refs. 64-66). The solar constant is 94% of the present level. This figure is reproduced from the Figure 1(a) of ref. 74 with permission from AGU.

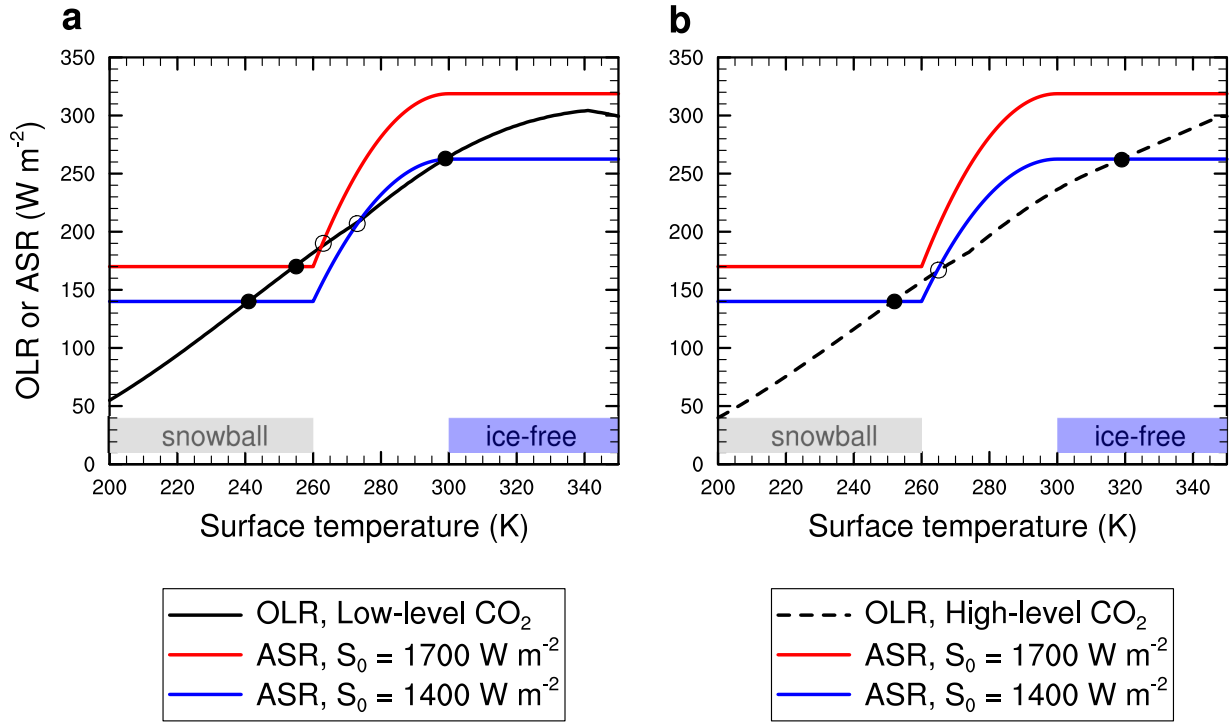

Fig. S9. Possible equilibrium climate states of a planet simulated by a 0D energy balance climate model. **a**, Outgoing longwave radiation (OLR) and absorbed stellar radiation (ASR) as a function of surface temperature for a low $CO_2$ concentration. **b**, Same as **(a)** but for a high $CO_2$ concentration (the OLR curve shifts downward). One equilibrium state is the point where the OLR curve and the ASR curve intersects; a filled cycle denotes a stable equilibrium state whereas an open cycle unstable. The stellar flux ($S_0$) is 1400 W m$^{-2}$ (blue line) and 1700 W m$^{-2}$ (red line). At high surface temperatures, the OLR curve asymptotes to a limiting value. Note that the ASR curves do not depend on the $CO_2$ concentration in this simple model.

**References in Supplementary Information:**